\documentclass[aps,prl,twocolumn]{revtex4-1}

\usepackage{graphicx}  
\graphicspath{ {images/} } 
\usepackage{latexsym}
\usepackage{slashed}
\usepackage{dcolumn}   
\usepackage{bm}        
\usepackage{amsmath}
\usepackage{amssymb}   
\usepackage{longtable}
\usepackage{float}
\newcommand\barparen[1]{\overset{(-)}{#1}}
\newcommand{\be}{\begin{equation}}
\newcommand{\ee}{\end{equation}}
\newcommand{\bea}{\begin{eqnarray}}
\newcommand{\eea}{\end{eqnarray}}
\newcommand{\bma}{\begin{matrix}}
\newcommand{\ema}{\end{matrix}}

\newcommand{\bml}{\begin{mathletters}}
\newcommand{\eml}{\end{mathletters}}
\newcommand{\bes}{\begin{subequations}}
\newcommand{\ees}{\end{subequations}}

\newcommand{\bi}{\begin{itemize}}
\newcommand{\ei}{\end{itemize}}
\newcommand{\gev}{~{\rm GeV}}
\newcommand{\tev}{~{\rm TeV}}
\newcommand{\mev}{~{\rm MeV}}

\begin{document}
\title{The Weak Eightfold Way: $SU(3)_W$ unification of the electroweak interactions}
\author{P. Q. Hung}
\email{pqh@virginia.edu}
\affiliation{Department of Physics, University of Virginia,
Charlottesville, VA 22904-4714, USA}

\date{\today}

\begin{abstract}
In a recent work, a successful prediction has been made for $\sin^2 \theta_W$ at an energy scale of O(TeV) based on the Dirac quantization condition of an electroweak monopole of the EW-$\nu_R$ model. The fact that such a prediction can be made has prompted the following question: Can $SU(2)$ be unified with $U(1)$ at O(TeV) scale since a prediction for $\sin^2 \theta_W$ necessarily relates the $U(1)$ coupling $g^{\prime}$ to the $SU(2)$ weak coupling $g$? It is shown in this manuscript that this can be accomplished by embedding $SU(2) \times U(1)$ into $SU(3)_W$ (The Weak Eightfold Way) with the following results: 1) The same prediction of the weak mixing angle is obtained; 2) The scalar representations of $SU(3)_W$ contain all those that are needed to build the the EW-$\nu_R$ model and, in particular, the real Higgs triplet used in the construction of the electroweak monopole.  3) Anomaly freedom requires the existence of mirror fermions present in the EW-$\nu_R$ model. 4) Vector-Like Quarks (VLQ) with unconventional electric charges are needed to complete the $SU(3)_W$ representations containing the right-handed up-quarks, with interesting experimental implications such as the prediction of doubly-charged hybrid mesons.
\end{abstract}

\maketitle

\section{Introduction}
Since the invention of the concept of the weak angle $\theta_W$ in the Standard Model (SM) and its subsequent experimental measurements, it has been the quest of particle theorists to find an explanation for its value. Since $\sin^{2} \theta_W = g^{\prime \, 2}/(g^{\prime \, 2} + g^{2})$ where $g^{\prime}$ and $g$ are the SM $U(1)$ and $SU(2)$ gauge couplings respectively, its prediction necessarily entails a relationship between $g^{\prime}$ and $g$. 

The traditional way to proceed is to embed the SM into a larger symmetry group in which $U(1)$ and $SU(2)$ are subgroups. The pioneering work was carried out \cite{G-G} where $SU(3)_c \times SU(2) \times U(1)$ was embedded in a Grand Unified gauge group $SU(5)$ and where  the prediction $\sin^{2} \theta_W = 3/8$ was made at the Grand Unified scale $M_{GUT}$. The $SU(5)$ model also predicted the decay of the proton. Renormalization of $\sin^{2} \theta_W=3/8$ down to the Z-mass, $M_Z$, gives a value that depends crucially on $M_{GUT}$. The initial calculations led to a proton lifetime which was ruled out by experiment. Further modifications of the initial proposal were made, including the introduction of supersymmetric versions and various scales between $M_{GUT}$ and $M_Z$, resulting in a larger proton lifetime and a value for the weak mixing angle $\sin^{2} \theta_W (M_Z) \approx 0.233$. It is not the purpose of the present manuscript to review the status of Grand Unified Theories (GUT) but to simply mention the most popular scenario involved in the prediction of $\sin^{2} \theta_W$.

The approach presented here will be very different from the GUT one in the sense that we adopt a "bottom-up" approach rather than the aforementioned "top-down" one. The principal motivation is the prediction of $\sin^2 \theta_W$ to be $1/4$ \cite{EHM} at an energy scale 2-3 TeV coming from the imposition of the Dirac quantization condition on the electroweak monopole of Ref.~\cite{pq_monopole}. This value of $1/4$ \cite{PUT} was evolved down at one loop to $\sim 0.231$ at $M_Z$. This rather surprising result in the absence of any unification could hint at a deeper phenomenon: the possible unification of $SU(2)$ and $U(1)$ at O(TeV) scale. The reason is simple: $\sin^2 \theta_W=1/4$ implies that $g^{\prime \, 2}=g^{2}/3$ since $\sin^{2} \theta_W = g^{\prime \, 2}/(g^{\prime \, 2} + g^{2})$. If such a unification were possible, what would be the simplest way to construct a "weak" unified group which is {\em anomaly-free} and which gives the particle content of the EW-$\nu_R$ model \cite{pqnur} used the construction of the non-sterile right-handed neutrinos and the electroweak monopole? It will turn out that by embedding $SU(2)_W \times U(1)_Y$ into $SU(3)_W$- the subscript $W$ refers to the fact that the weak gauge bosons interact with SM and mirror fermions- and by requiring $SU(3)_W$ to be anomaly-free, the existence of mirror fermions- the key ingredient of the EW-$\nu_R$ model- arises {\em naturally}. (Mirror fermions were introduced to provide a model for EW-scale, non-sterile right-handed neutrinos.)

The plan of this paper is as follows. First, a brief review of the EW-$\nu_R$ model \cite{pqnur} is presented: its motivation, its particle content, and finally the construction of the electroweak monopole. This is followed by the description of the derivation of $\sin^2 \theta_W=1/4$  at an energy scale of O(few TeV) corresponding to the mass of the electroweak monopole. Second, we will show the steps used in the construction of the unifying group $SU(3)_W \supset SU(2)_W \times U(1)_Y$. In particular, it will be shown that the requirement of anomaly-freedom for $SU(3)_W$ leads automatically to the introduction of mirror fermions as the only option. Mirror fermions are $SU(2)_W$ right-handed doublets and left-handed singlets, a key ingredient of the EW-$\nu_R$ model.  The scalar sector is found to be quite interesting in its own right. Finally, it turns out that Vector-Like Quarks (VLQ) with unconventional electric charges are needed to complete the $SU(3)_W$ representations containing the right-handed up-quarks, with interesting experimental implications such as the prediction of doubly-charged hybrid mesons containing one VLQ and one ordinary quark. Gauge interactions are mediated by heavy TeV-scale gauge bosons linking members of a $SU(2)_W$ doublet (e.g. $(\nu_L, e_L)$) to a singlet (e.g. $e_R$ ($e^{c}_L$)). Signals of such interactions can be searched for at the LHC and in precision experiments such as the left-right asymmetry in polarized electron scattering on unpolarized proton at Jefferson Laboratory.

\section{The road to the $SU(3)_W$ unification of $SU(2)_W \times U(1)_Y$} 
\label{summary}

\subsection{The EW-$\nu_R$ model: a summary}

The EW-$\nu_R$ model's \cite{pqnur,pqnur2} main aim was the construction of a seesaw mechanism in which right-handed neutrinos acquire masses proportional to the electroweak scale $\Lambda_{EW}= 246 \gev$ and are {\em non-sterile}. The reasons for doing so are three-fold: 1) The origin of neutrino masses being, without doubt, one of the most important problems in Physics begs for an urgent question: At what energy scale(s), does it show up? Could it be at around the electroweak scale?; 2) There is no known physical principle that requires right-handed neutrinos to be singlets under $SU(2)_W \times U(1)_Y$; 3) It is then attractive to entertain the idea that right-handed neutrinos are actually {\em non-singlets} under $SU(2)_W \times U(1)_Y$ and acquire masses through the spontaneous breakdown of the electroweak gauge group. The implication of this idea is the possibility of directly testing the seesaw mechanism at colliders such as the LHC by searching for its progenitors: the non-sterile right-handed neutrinos.

The construction of the EW-$\nu_R$ model \cite{pqnur} is, in fact, quite simple. In what follows, we shall list the particle content and the corresponding interactions, in preparation for the unification discussion.

\bi
\item {\bf Gauge group}: The same gauge group as in the SM i.e. $SU(3)_C \times SU(2)_W \times U(1)_Y$.

\item {\bf Fermions}: The simplest way to have non-sterile right-handed neutrinos is to put them in $SU(2)_W$ doublets along with mirror right-handed charged leptons. Anomaly freedom dictates that mirror quarks should also be included. We have

SM: $l_L = \left(
	  \begin{array}{c}
	   \nu_L \\
	   e_L \\
	  \end{array}
	 \right)$; $q_L = \left(
	  	 \begin{array}{c}
	   	  u_L \\
	     	  d_L \\
	  	\end{array}
	 	\right)$; $e_R; \ u_R, \ d_R$;

Mirror: $l_R^M = \left(
	  \begin{array}{c}
	   \nu_R \\
	   e_R^M \\
	  \end{array}
	 \right)$; $q_R^M = \left(
	  	 \begin{array}{c}
	   	  u_R^M \\
	     	  d_R^M \\
	  	\end{array}
	 	\right)$; $e_L^M; \ u_L^M, \ d_L^M$.
		
\item {\bf Scalars}: 

\underline{Doublet Higgs fields} \cite{pqnur,pqcpnew} (similar to 2HDM): $\Phi^{SM}_{1} (Y/2=-1/2)$, $\Phi^{SM}_{2} (Y/2=+1/2)$ coupled to SM fermions and $\Phi^{M}_{1} (Y/2=-1/2)$, $\Phi^{M}_{2} (Y/2=+1/2)$ coupled to mirror fermions with $\langle \Phi^{SM}_{1} \rangle=(v_1/\sqrt{2},0)$,  $\langle \Phi^{SM}_{2} \rangle=(0,v_2/\sqrt{2})$ and $\langle \Phi^{M}_{1} \rangle=(v^M_1/\sqrt{2},0)$, $\langle \Phi^{M}_{2} \rangle=(0,v^M_2/\sqrt{2})$ .

\underline{Triplet Higgs fields}: One complex $\tilde{\chi}=(\chi^0,\chi^+,\chi^{++})$ and one real $\xi (Y/2 = 0)= (\xi^+, \xi^0, \xi^-)$ which can be put in a 3x3 matrix of global $SU(2)_L \times SU(2)_R$

$\chi = \left( \begin{array}{ccc}
\chi^{0} &\xi^{+}& \chi^{++} \\
\chi^{-} &\xi^{0}&\chi^{+} \\
\chi^{--}&\xi^{-}& \chi^{0*}
\end{array} \right)$

with $\langle \chi^0 \rangle = \langle \xi^0 \rangle=v_M$ in order to preserve Custodial Symmetry (that guarantees $M_{W}^2 = M_{Z}^2 \cos^{2}\theta_W$ at tree level). Here $\sum_{i=1,2} (v_i^2 + v^{M,2}_{i}) + 8 v_M^2 = (246 GeV)^2$.	 

An important point is in order at this point. As mentioned above, the {\em real} triplet $\xi$ was introduced in order to maintain the custodial symmetry, once the complex triplet $\tilde{\chi}$ was used to provide electroweak-scale Majorana mass to right-handed neutrinos. It is this real triplet that provide the monopole solution to the  EW-$\nu_R$ model. 

\underline{Complex singlet Higgs fields}:
$\phi_S$ with $\langle \phi_S \rangle =v_S \ll v_M$ (to be explained below): Important scalars connecting the SM and Mirror worlds. 
\item {\bf Yukawa interactions}: In what follows are the relevant Yukawa interactions responsible for the seesaw mechanism.
\begin{itemize}
\item Majorana mass terms:

The complex Higgs triplet $\tilde{\chi}$ can be written as
\be
\label{tildechi}
\tilde{\chi}=\frac{1}{\sqrt{2}} \vec{\tau}.\vec{\chi}=\left( \begin{array}{cc}
\frac{1}{\sqrt{2}} \chi^{+}&\chi^{++} \\
\chi^{0}&-\frac{1}{\sqrt{2}} \chi^{-}
\end{array} \right) \,
\ee
and, noticing that $l^{M,T}_R  \sigma_2 l^M_R$ is the fermion bilinear which contains the $\nu_R$ Majorana mass term, one obtains
\be
\label{Majorana}
L_M =  g_M \, l^{M,T}_R  \sigma_2 (\tau_2 \ \tilde{\chi}) \ l^M_R,
\ee
giving $M_R = g_M v_M $ $\Rightarrow$ $M_Z/2<M_R < O(\Lambda_{EW} \sim 246 GeV)$. 
\item Dirac mass terms:
\be
\label{Dirac}
{\mathcal L}_S = - g_{Sl} \,\bar{l}_L \ \phi_S \ l_R^M + {\rm H.c.},
\ee
giving $m_D = g_{Sl} \ v_S $. 

\item Light neutrino mass: $|m_{\nu}| \sim m_{D}^2/M_R = (g^{2}_{Sl}/g_M) (v^{2}_S/v_M)$; Right-handed Majorana mass: $M_R$. 

\item Charged fermion masses:

 \bea
 \label{massq}
 {\cal L}_{mass,q}& = & g_u \bar{q}_L  \Phi^{SM}_{1} u_R + g_d \bar{q}_L  \Phi^{SM}_{2} d_R  \nonumber \\
 &&+ g^{M}_{u} \bar{q}^M_R  \Phi^{M}_{1}  u^M_L+ g^{M}_{d} \bar{q}^M_R  \Phi^{M}_{2}  d^M_L  \nonumber \\
 &&+H.c. \,,
 \eea
\be
\label{massl}
 {\cal L}_{mass,l} =  g_{l} \bar{l}_L  \Phi^{SM}_{2} e_R + g^{M}_{l} \bar{l}^M_R  \Phi^{M}_{2}  e^M_L + H.c.
\ee

\end{itemize}
\item {\bf Global symmetries}:
The EW-$\nu_R$ model contains a global symmetry group $U(1)_{SM} \times U(1)_{MF}$ proposed to ensure a proper form of seesaw mechanism of the type shown in Eq.~(\ref{Majorana}), i.e. a term such as $g_M \, l^{T}_L  \sigma_2 \tau_2 \ \tilde{\chi} \ l_L,$ is forbidden by the symmetry.  We have
\begin{eqnarray*}
\label{SMtransformation}
			U(1)_{SM}
			&:&( l_L, q_L) \;\rightarrow e^{-\imath \alpha_{SM}} \;(l_L, q_L)\,,\\
			&& (e_R,u_R, d_R) \;\rightarrow e^{\imath \alpha_{SM}} \; (e_R,u_R, d_R) \,, \\
			&& \Phi^{SM}_{1,2} \rightarrow  e^{-2\imath \alpha_{SM}} \Phi^{SM}_{1,2} \,.
\end{eqnarray*} 
\begin{eqnarray*}
\label{MFtransformation}
			U(1)_{MF}
			&:& (l_R^M,q_R^M )\;\rightarrow e^{\imath \alpha_{MF}} \; (l_R^M,q_R^M)\,,\\
			&& (e_L^M,u_L^M, d_L^M) \;\rightarrow e^{-\imath \alpha_{MF}} \; (e_L^M,u_L^M, d_L^M) \,, \\
			&& \Phi^{MF}_{1,2} \rightarrow  e^{2\imath \alpha_{MF}} \Phi^{MF}_{1,2} \,,
\end{eqnarray*} 
\be
\label{singlet}
\phi_S \rightarrow e^{-\imath(\alpha_{SM} + \alpha_{MF})} \phi_S \,,
\ee
and
\be
\label{tripletchi}
\chi \rightarrow e^{-2\imath \alpha_{MF}} \chi \,.
\ee
\ei

The above transformations came from the  EW-$\nu_R$ model. However, as we shall see below, one can modify the transformations on $u_R$, $u^{M}_R$, $\Phi^{SM}_1$ and $\Phi^{MF}_{1}$ without changing the physics of the model. In fact, the $SU(3)_W$ requires such a change.
\subsection{The electroweak monopole and $\sin^{2} \theta_W$}

Hidden in the EW-$\nu_R$ model is a "topological" surprise: the existence of an electroweak magnetic monopole \cite{pq_monopole}. The SM which contains only a Higgs doublet can be shown to contain no topologically stable, finite energy monopole. The Cho-Maison monopole \cite{cho-maison} can however exist in the SM but with a catch: it does not have a finite energy solution unless some unknown physics beyond the SM is invoked resulting in the modification of the $U(1)_Y$ gauge kinetic term \cite{cho-maison2}. If we wish to keep the same gauge structure of the SM with no modification, the only possible option is to endow the Higgs content of the SM gauge group with a real Higgs triplet, which is not the case with the minimal SM. In what follows is a brief summary of some of the essential features in the construction of a topologically stable, finite energy monopole.

In searching for models which could give rise to a topologically stable monopole, the usual route is to map a spatial 3-dimensional sphere $S^2$ to the space of the vacuum manifold $S_{vac}$ of the model. In the parlance of homotopy theory, this amounts to see if the second homotopy group $\pi_{2}(S_{vac}))$ is {\em non-trivial}. Furthermore, homotopy theory tells us that $\pi_{2}(S^n)=0$ for $n>2$ and $\pi_{2}(S^2)=Z$, where $Z=0,1,2,..$. The vacuum manifold of the SM with only a complex Higgs doublet (four independent degrees of freedom) is represented by $\phi_1^2 + \phi_2^2 + \phi_3^2 + \phi_4^2 = v^2$. This is a 3-sphere $S^3$ and $\pi_{2}(S^3)=0$. The minimal SM cannot accommodate a topologically stable monopole. 

A {\em real} Higgs triplet has a vacuum manifold $\xi_0^2+\xi_1^2+\xi_2^2=v_M^2$ corresponding to $S^2$ and, in consequence, the EW-$\nu_R$ model can accommodate a topologically stable monopole. (Let us notice one more time that this real Higgs triplet was introduced in order to maintain the experimentally successful custodial symmetry.) 
The vacuum manifold of the model is $S_{vac} = S^{2} \times S^{5} \times \sum_{i=1,2} S^{3}_{SMi} \times \sum_{i=1,2} S^{3}_{Mi} $, where $S^3$ and $S^5$ denote the vacuum manifolds of the Higgs doublets and complex triplet respectively. Furthermore, $\pi_2(S_{vac})=\pi_2(S^2)  \oplus \pi_2(S^5)  \oplus_{i=1,2} \pi_2(S^{3}_{SMi,Mi}) =  \pi_2(S^2) = Z$, justifying the topological stability of the electroweak monopole. 
It is also a monopole with {\em finite energy} since it is of the  't Hooft-Polyakov type \cite{thooft}. In fact, the 't Hooft-Polyakov ans$\ddot{a}$tz for $\xi$ is given by $\xi^a = \frac{r^a}{g r^2} H(v_M g r)$ with $\xi^a \rightarrow v_M r^a/r$ as $r \rightarrow \infty$. As an example, one can take the triplet self-coupling $\lambda_3 \sim 1$ and one can see that $H(v_M g r) \rightarrow v_M g r$ rapidly for $r \sim 2/(gv_M)$. The electroweak monopole has a core of radius $R_c \sim (gv_M)^{-1}$ and a mass $M=\frac{4\pi v_M}{g} f(\lambda/g^2)$. At this point, it is worth emphasizing the deep connection, in the EW-$\nu_R$ model, between properties of the monopole such as its mass and size, and a key property of the non-sterile right-handed neutrinos which is their mass: {\em both are proportional to the VEV of the triplets}, namely $v_M$. The search for one has experimental consequences on the other.

In what follows, we will present a summary of two arguments leading to the prediction of $\sin^{2} \theta_W$: the topological quantization and the Dirac quantization conditions. 

1) Since $S^2$ corresponds to the vacuum manifold of the real triplet $\xi$, the topological quantization involves the $SU(2)_W$ coupling $g$ and the field strength $W_{3}^{\mu \nu}=\partial^{\mu} W_{3}^{\nu}-\partial^{\nu} W_{3}^{\mu} + \frac{1}{v_{M}^{3} g} \varepsilon_{abc} \xi^{a} \partial^{\mu} \xi^{b} \partial^{\nu} \xi^{c} $. A (conserved) topologically current can be constructed: $k_\mu = \frac{1}{2} \epsilon_{\mu \nu \sigma \rho} \partial^{\nu} W_{3}^{\sigma \rho}$, and the "magnetic" charge can be computed from $g_{M}=\int d^{3} x k_{0}$. A straightforward calculation gives the topological quantization condition: $\frac{g g_{M}}{4\pi}=n$. As it has been done in the literature, one can bring it to a form similar to that of the Dirac quantization condition by the following re-definttion: $(g,g_{M})/\sqrt{4\pi} \rightarrow (g,g_{M})$ , giving the following topological quantization condition (TQC):
\be
\label{topological}
g g_M = n \,.
\ee
From TQC, the smallest magnetic charge corresponding to $n=1$ is $g_M=1/g$.
As emphasized in Ref.~(\cite{pq_monopole}), since $\xi$ which carries zero $U(1)_Y$ quantum number, its VEV would induce $SU(2)_W \times U(1)_Y \rightarrow U(1)_W \times U(1)_Y$. The magnetic field strength $W^{3}_{ij} = \frac{\epsilon_{ijk} \hat{\vec{x}}^k}{g r^2}$ and the long-range magnetic field $B_{i} = -\frac{1}{2} \epsilon_{ijk} W^{3,jk}=\frac{1}{g r^2} \hat{r}_{i}$ are associated with that symmetry breaking step. The complex triplet $\tilde{\chi}$ and the doublet Higgs fields spontaneously break $SU(2)_W \times U(1)_Y$ down to $U(1)_{em}$. With $W^{3}_{ij} =  \cos \theta_W Z_{ij} + \sin \theta_W F_{ij}$ and $g=e/ \sin \theta_W$ , the true long-range magnetic field comes from $F_{ij}$ and takes the form 
 \be
 \label{magfield}
B_{i}  = \frac{g_M}{ r^2} \hat{r}_{i}  = \frac{\sin \theta_W}{e r^2} \hat{r}_{i} \,, 
\ee
for the minimal case $n=1$.

2) The requirement that the wave function of an electron in the presence of a monopole of charge $g_M$ gives rise to the Dirac Quantization Condition (DQC)
\be
\label{DQC}
e g_M = \frac{m}{2} \,.
\ee
When we combine DQC with TQC, namely $g_M=\frac{n \, \sin \theta_W }{e}$ and take $n=1$, we obtain \cite{EHM}
\be
\label{weak}
\sin \theta_W = \frac{m}{2} \,.
\ee
The only possible value allowed for $m$ is $m=1$ giving the prediction
\be
\label{prediction}
\sin^{2} \theta_W= \frac{1}{4} \,.
\ee
The above prediction by \cite{EHM} for $\sin^{2} \theta_W$ is understood to be the value at the mass scale of the monopole. The monopole mass, $M_{M}=\frac{4\pi v_M}{g} f(\lambda/g^2) \sim 889 \gev - 2.993 \tev$. (The value of $v_M$ is constrained from below by the Z-width (only three light neutrinos) i.e. $v_M > M_Z/2 \sim 45.5 \gev$ and from above by $\sum_{i=1,2} (v_i^2 + v^{M,2}_{i}) + 8 v_M^2 = (246 \gev)^2$. Taking the lowest value corresponding to $v_M \sim 45.5 \gev$ and $f=1$ and the largest value corresponding to $v_M \sim 87 \gev$ and $f=1.78$, one obtains that range of monopole mass.) Leading-logarithmic, one-loop calculations of the renormalization of $\sin^{2} \theta_W$ from $M_M$ down to the Z-boson mass $M_Z$ led to the following estimates: $\sin^{2} \theta_W (M_Z) =0.231-0.232$ for $M_M = 2.3 \tev - 3 \tev$ and for various choices of spectra. Last but not least, if we identify the topological number $n$ from TQC with the integer $m$ from DQC, we obtain the result (\ref{prediction}) for {\em any} magnetic charge and not just $g_M=1/g$. 

Since ${\sin}^2\theta_W(M_M^2) = g'^2/[g^2+g'^2]=1/4$ giving $\alpha^\prime =(1/3) \alpha_2$ ($g$ and $g^{\prime}$ are the $SU(2)_W$ and $U(1)_Y$ couplings respectively), the obvious question is whether or not there could be a unification of $SU(2)_W$ with $U(1)_Y$ at $M_M$ because of the relationship between $g$ and $g^{\prime}$. The answer is {\em yes} and it is the subject of the next section.

\section{The $SU(3)_W$ unification of $SU(2)_W$ and $U(1)_Y$}

The path to finding the embedding of $SU(2)_W \times U(1)_Y$ is surprisingly simple. The requirements for $G \supset SU(2)_W \times U(1)_Y$ are the usual ones, namely $G$ should be anomaly-free and it should contain the particle content at least of the SM model. 

The next question one would like to ask is the physical meaning of the $SU(2)_W \times U(1)_Y$ unification. We have already seen long ago the embedding of $SU(3)_c \times SU(2)_W \times U(1)_Y$ into $SU(5)$ as an example of quark-lepton unification since that also involves the color gauge group. The unification in question here is simply one between a weak doublet and a weak singlet as we shall see below.

\subsection{The choice of $SU(3)_W$ as an unification group}
The most obvious choice for $G$ is $SU(3)_W$ which can be spontaneously broken down to $SU(2)_W \times U(1)_Y$ by an adjoint Higgs field at a unification scale $M_U$. The generators of $SU(3)_W$ are $T_a = \lambda_a / 2$ where $\lambda_a$ are the usual Gell-Mann matrices. The two diagonal generators are $T_3 =\lambda_3/2$ (now part of $SU(2)_W$) and $\lambda_8/2$ which is related to the weak hypercharge $Y/2$. Now, let us notice that $\alpha^\prime =(1/3) \alpha_2$ is equivalent to $g^{\prime} = g/\sqrt{3}$ at $M_U$. It is $g_1= \sqrt{3} g^{\prime}$ that is equal to $g$ at $M_U$. It is straightforward to see this by looking at the covariant derivatives for $SU(3)_W$, namely ($a=1,..,8$; $i=1,2,3$)
\be
\label{covariant1}
D_{\mu}= \partial_{\mu} + \imath g_U (\frac{\lambda^a}{2}) A^{a}_{\mu} \,,
\ee
and for $SU(2)_W \times U(1)_Y$,
\be
\label{covariant2}
D_{\mu}= \partial_{\mu} + \imath g (\frac{\tau^i}{2}) W^{i}_{\mu} + \imath g^{\prime} (\frac{Y}{2}) B_{\mu}  \,.
\ee
Identifying $A^{8}_{\mu}$ with $B_{\mu}$, one immediately obtains
\be
\label{hyper}
g_{U} (\frac{\lambda_8}{2})= g^{\prime} (\frac{Y}{2}) \,.
\ee
At the unification scale $M_U$, $g=-\sqrt{3} g^{\prime}=g_U$ and with $\lambda_{8}/2=diag(1,1,-2)/2\sqrt{3}$, one readily obtains
\be
\label{hypercharge2}
\pm \frac{Y}{2}= diag(\pm \frac{1}{2}, \pm \frac{1}{2}, \mp1) \,.
\ee

One could also turn the argument around and requires that the hypercharge operator is given by Eq.~(\ref{hypercharge2}) as it would be the case with the SM. One would then obtain $g^{\prime}= g/\sqrt{3}$ and get ${\sin}^2\theta_W(M_M^2) = 1/4$. The Dirac Quantization Condition discussed above is {\em automatically satisfied}. 

The hypercharge operator (\ref{hypercharge2}) will be at the core of the search for the right fermion representations of the model.

\subsection{The weak "Eightfold way": $SU(3)_W \rightarrow SU(2)_W \times U(1)_Y$}
I) {\bf $SU(3)_W$ symmetry breaking}:

The eight gauge bosons of $SU(3)_W$ decompose under $SU(2)_W \times U(1)_Y$ as ${\bf G}_{\mu}= (W^{\pm}_{\mu}, W^{3}_{\mu}) + B_{\mu} + (X^{+}_{\mu}, X^{0}_{\mu})+(X^{-}_{\mu},\bar{X}^{0}_{\mu})$. Here $W^{\pm}_{\mu}, W^{3}_{\mu}$ and $B_{\mu}$ are the usual gauge fields of  $SU(2)_W \times U(1)_Y$. This can be put in a matrix form as
\be
\label{gauge}
(\frac{\lambda^a}{2}) A^{a}_{\mu} =  \left(
	  \begin{array}{ccc}
	  \frac{W_{3\mu}}{2} +\frac{B_\mu}{2\sqrt{3}}, \, W^{+}_{\mu}, \, \frac{X^{+}_{\mu}}{\sqrt{2}} \\
	  W^{-}_{\mu}, \, -\frac{W_{3\mu}}{2} +\frac{B_\mu}{2\sqrt{3}}, \, \frac{X^{0}_{\mu}}{\sqrt{2}} \\
	  \frac{X^{-}_{\mu}}{\sqrt{2}}, \, \frac{\bar{X}^{0}_{\mu}}{\sqrt{2}}, \, -\frac{B_\mu}{\sqrt{3}} \\
	    \end{array}
	 \right) \,.
\ee

In the breaking $SU(3)_W \rightarrow SU(2)_W \times U(1)_Y$, the $ (X^{+}_{\mu}, X^{0}_{\mu})+(X^{-}_{\mu},\bar{X}^{0}_{\mu})$ gauge fields obtain a mass proportional to the breaking scale. 

The appropriate Higgs field for that symmetry breaking is an octet $\Phi_{8}$ .The particle content of $\Phi_{8}$ under $SU(2)_W \times U(1)_Y$ is $\Phi_8 = (3,0)+(2,1/2)+ (0,0)$ and written explicitly as $\Phi_8 = (\xi^{+}, \xi^{0}, \xi^{-}) + (\phi_X^{+},\phi^{0}_X) + (\phi_X^{-},\bar{\phi}^{0}_X) + \eta_8$, or in matrix form as 
\be
\label{octethiggs}
\Phi_8 =  \left(
	  \begin{array}{ccc}
	  \frac{\xi^{0}}{2} +\frac{\eta_8}{2\sqrt{3}}, \, \xi^{+}, \, \frac{\phi^{+}_{X}}{\sqrt{2}} \\
	  \xi^{-}, \, -\frac{\xi^{0}}{2} +\frac{\eta_8}{2\sqrt{3}}, \, \frac{\phi^{0}_{X}}{\sqrt{2}} \\
	  \frac{\phi^{-}_{X}}{\sqrt{2}}, \, \frac{\bar{\phi}^{0}_{X}}{\sqrt{2}}, \, -\frac{\eta_8}{\sqrt{3}} \\
	    \end{array}
	 \right) \,.
\ee
whose VEV takes the form
\be
\label{octetvev}
\langle \Phi_8 \rangle = v_8 \, diag(1,1,-2) \,,
\ee
where
\be
\label{vev}
v_8=\frac{\langle \eta_8 \rangle}{2 \sqrt{3}} \,.
\ee
First, one can identify, after symmetry breaking, $(\phi_X^{+},\phi^{0}_X) + (\phi_X^{-},\bar{\phi}^{0}_X)$ as the Nambu-Goldstone bosons which are absorbed by the gauge fields $(X^{+}_{\mu}, X^{0}_{\mu})+(X^{-}_{\mu},\bar{X}^{0}_{\mu})$ to get masses of order $O(v_8)$. After symmetry breaking, the remaining physical Higgs fields of $\Phi_8$ are the following ones: a {\em real} triplet $\xi= (\xi^{+}, \xi^{0}, \xi^{-})$ 
and the {\em real} electroweak singlet $\eta_8$ which is different from the {\em complex} singlet $\phi_S$. Although the {\em real} triplet $\xi= (\xi^{+}, \xi^{0}, \xi^{-})$ has the right quantum number as the one of the EW-$\nu_R$ model whose VEV ensures the custodial symmetry and gives rise to the electroweak monopole, it however appears not to be the correct one because it would acquire a physical mass after this first step of symmetry breaking. Also, as we show below, there are two other {\em real} triplets and the one that we need for the aforementioned custodial symmetry could be a combination of the three real triplets discussed here. This, however, will be investigated further in a separate paper.
\\
\\
II) {\bf Left-handed fermion representations of $SU(3)_W$}:
\\
\\

We first ask which $SU(3)_W$ representations one can choose to fit all the SM fermion degrees of freedom. The next requirement will be that of anomaly freedom.

The basic SM fermion degrees of freedom are summarized as follows, for each generation: $(\nu_L, e_L)$, $e_R$  for the leptons and $(d_L, u_L)$, $u_R, d_R$ for the quarks. Following the practice of the construction of grand unified models, we group these SM degrees of freedom in left-handed multiplets using $\psi^{c}_{L}=C \bar{\psi}_{R}^{T}$. To see what representations of $SU(3)_W$ are needed to accommodate those fermions, we need to set up an orthonormal basis: $T_{3W}$ and $\frac{Y}{2}$ and classify representations which have simultaneous eigenvalues of those operators and $T_{W}^2$. (These steps are similar to the ones used in the quintessential $SU(3)$ flavor symmetry.) $SU(3)_W$ representations are denoted by a pair of positive integers $(m,n)$ with dimension $(m+1)(n+1)(m+n+2)/2$. The eigenvalues of $T_W$ ($T_{3W}=-T_W,..,T_W$) and $Y/2$ are
\be
\label{T}
T_W=(p+q)/2 \,,
\ee
and
\be
\label{Yq}
\frac{Y_q}{2}=\frac{p}{2}-\frac{q}{2}+\frac{1}{3}(n-m) \,,
\ee
for quarks, with
\be
\label{pq}
0 \leq p \leq m; 0 \leq q \leq n \,. 
\ee
For leptons, $Y_l/2$ is given by Eq.~(\ref{hypercharge2}).
\\
\\
a) {\bf Leptons}: 
\\
\\
We have $(\nu_L, e_L)$, $e_R$ $\rightarrow$ $(\nu_L, e_L)$, $e^{c}_L$. This suggests a triplet: $(1,0)$ or $(0,1)$.

First, $\frac{Y_l}{2}(e^{c}_L)=1$. From Eq.~({\ref{hypercharge2}), for the doublet, one obtains $\frac{Y_l}{2}=-1/2$. This corresponds to $\bar{3}_l=(0,1)$. The graphical representation is shown in Fig.~(\ref{e3}).
\begin{figure}
\begin{center}
\includegraphics[width=0.2\textwidth]{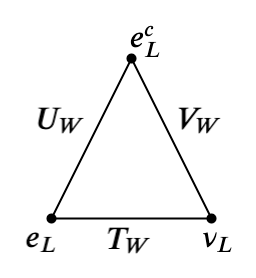}
\end{center}
\caption{Representation $\bar{3}_l$}
\label{e3}
\end{figure}

In analogy with the old Eightfold Way, $T$, $U$ and $V$ are defined by
\be
\label{TUV}
T^{\pm}_W=\frac{\tau_1 \pm \imath \tau_2}{2};\, U^{\pm}_W=\frac{\lambda_6 \pm \imath \lambda_7}{2};\, V^{\pm}_W=\frac{\lambda_4 \pm \imath \lambda_5}{2} \,.
\ee
With
\be
\label{gaugeWX}
W^{\pm}_{\mu}=\frac{W^{1}_{\mu} \mp \imath W^{2}_{\mu}}{\sqrt{2}}; \, X^{\pm}_{\mu} = \frac{A^{4}_{\mu} \mp \imath A^{5}_{\mu} }{\sqrt{2}}; \, \barparen{X^{0}}_{\mu} = \frac{A^{6}_{\mu} \mp \imath A^{7}_{\mu} }{\sqrt{2}}
\ee
The above generators are coupled to the gauge bosons as follows:
\be
\label{TUVgauge}
\frac{1}{\sqrt{2}}(T^{\pm}_W \, W^{\pm}_{\mu} + V^{\pm}_W \, X^{\pm}_{\mu} + U^{\pm}_W \, \barparen{X^{0}}_{\mu} ) \,.
\ee

Notice that, since $Q=T_{3W}+ Y/2$, $TrQ=0$ as it should. We shall denote this leptonic representation as $\bar{{\bf 3}}^{l}_{L}$.
\be
\label{lepton}
 \bar{{\bf 3}}^{l}_{L}= \left(
	  \begin{array}{c}
	   \nu_L (0)\\
	   e_L  (-1)\\
	   e^{c}_L (+1)\\
	  \end{array}
	 \right) \,.
\ee

Under $SU(3)_c \times SU(3)_W$, $ \bar{{\bf 3}}^{l}_{L} = (1, \bar{3})$.
\\
\\
b) {\bf Down quarks}:
\\
\\
We have $(d_L, u_L)$, $d_R$ $\rightarrow$ $(d_L, u_L)$, $d^{c}_L$. Now, $\frac{Y_d}{2}(d^{c}_L)=1/3$. From Eq.~(\ref{Yq}), one obtains the eigenvalues $1/3$ and $-1/6$. This again suggests the $(0,1)$ representation where the doublet is represented by $(u^{*}_L (-2/3) ,\, d^{*}_L (1/3) )$. The graphical representation is shown in Fig.~(\ref{d3}).
\begin{figure}
\begin{center}
\includegraphics[width=0.2\textwidth]{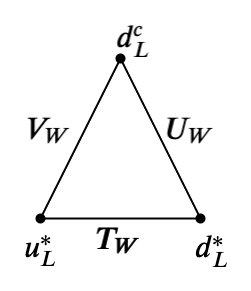}
\end{center} 
\caption{Representation }
\label{d3}
\end{figure}

The appropriate transformation that one should use is $\imath \tau_2 q^{*}_L=(d^{*}(1/3), -u^{*}(-2/3))$. Our "down-quark" representation takes the form
\be
\label{down}
 \bar{{\bf 3}}^{d}_{L}= \left(
	  \begin{array}{c}
	   d^{*}_L (1/3) \\
	   -u^{*}_L (-2/3) \\
	   d^{c}_L (1/3)\\
	  \end{array}
	 \right) \,.
\ee

Since $d^{c}_L$ is a member of $ \bar{{\bf 3}}^{d}_{L}$ of $SU(3)_W$ and it is a {\bf $\bar{3}$} under $SU(3)_c$, it implies that  $\bar{{\bf 3}}^{d}_{L}$ transforms as {\bf $\bar{3}$} under $SU(3)_c$ as well. Under $SU(3)_c \times SU(3)_W$, $ \bar{{\bf 3}}^{d}_{L}= (\bar{3}, \bar{3})$.
\\
\\
c) {\bf Up quarks}
\\
\\
Here we have the $SU(2)_W$ singlet $u_R$ which becomes $u^{c}_L (-2/3)$. From Eq.~(\ref{Yq}), one deduces that $m=2$ and $n=0$. The representation $(2,0)$ corresponds to a {\em sextet} of $SU(3)_W$. The $Y/2$ eigenvalues are as follows: $-2/3$, $-1/6$ and $1/3$. $-1/6$ corresponds to a doublet {\em different} from $(d_L, u_L)$ since the usual quark doublet has already been included in the "down" representation above, and $1/3$ which gives  $p=2$ and $q=0$, and hence $T_W=1$ corresponds to a triplet with unconventional electric charges as illustrated in Fig.~(\ref{unew}).
\begin{figure}
\begin{center}
\includegraphics[width=0.35\textwidth]{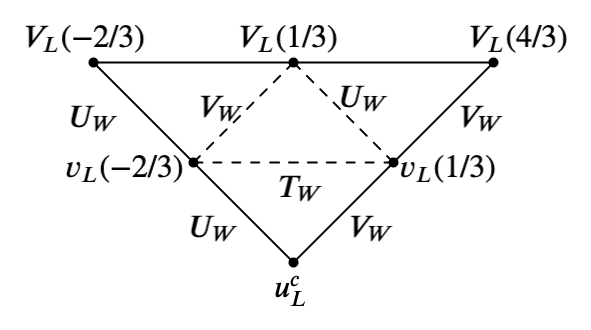}
\end{center} 
\caption{Representation ${\bf 6}_L$}
\label{unew}
\end{figure}

With $u^{c}_L (-2/3)$, we are forced to introduce a new $SU(2)_W$ doublet, $(\upsilon_L(-2/3), \upsilon_L(1/3))$, and a triplet of quarks with unconventional electric charges $(V_L (-2/3), V_L (1/3), V_L (4/3) )$. This sextet will be denoted by ${\bf 6}_L$. This unusual situation is reminiscent of the $SU(5)$ case where $q_L$ and $u^{c}_L$ belong to the 10-representation and $d^{c}_L$ belongs to the $\bar{5}$ representation. There are two obvious reasons why one cannot stop here: 1) these extra quarks would be massless since they do not have their right-handed partners (or equivalently $f^{c}_L$); 2) as we shall see below, $SU(3)_W$ would not be anomaly-free with only these representations. 

Under $SU(3)_c \times SU(3)_W$, ${\bf 6}_L= (\bar{3}, 6)$. 
\\
\\
III) {\bf Anomaly freedom and the need for mirror fermions}:
\\
\\
So far, following the logical rules for embedding $SU(2)_W \times U(1)_Y$ into $SU(3)_W$, we have managed to fit the SM quarks and leptons into three $SU(3)_W$ representations: $\bar{{\bf 3}}^{l}_{L}$. $\bar{{\bf 3}}^{d}_{L}$ and ${\bf 6}_L$ for each family. As argued above, we are forced to introduce a triplet of quarks with unconventional electric charges. Before addressing the issue of fermion masses in the model, the next step in our construction is to study the requirement of anomaly freedom.

The anomaly coefficients $A(R)$ for a representation $R$ of a gauge group $G$ can be calculated from
\be
\label{anomaly}
Tr[\{T^{R}_a,T^{R}_b\} T^{R}_c]= d_{abc} A(R) \,.
\ee
For $SU(N)$, the fundamental representation $F$ and its conjugate $\bar{F}$ have the following coefficients: $A(F)=1$ and $A(\bar{F})=-1$. The symmetric representation $(2,0)$ with dimension $N(N+1)/2$ has a coefficient $A{(2,0)}=N+4$. In our case of $SU(3)_W$, we have $A(\bar{{\bf 3}}^{l}_{L})= A(\bar{{\bf 3}}^{d}_{L})=-1$ and $A({\bf 6}_L)=7$. Taking into account the color factor, the total anomaly is $A_{tot}=-1-3+21= 17$. It is obvious that $\bar{{\bf 3}}^{l}_{L}$. $\bar{{\bf 3}}^{d}_{L}$ and ${\bf 6}_L$ are not sufficient to make $SU(3)_W$ {\em anomaly-free}.

At this point, it is worth to point out that it is not feasible to search for anomaly coefficients from {\em higher dimensional representations} for two reasons: 1) They necessarily {\em do not contain SM fermion degrees of freedom} and are unnecessarily complicated; 2) Their anomaly coefficients are positive, unless one looks at their conjugates. This is unlike the case with $SU(5)$ where all the SM degrees of freedom can fit snugly into a $10$ and $\bar{5}$ representations and $A(10)=1$ and $A(\bar{5})=-1$.

The simplest option to have an anomaly-free $SU(3)_W$ is to have representations with the {\em same dimensionality} but with {\em opposite anomaly coefficients}. This is where mirror fermions come in. The anomalies of the left-handed representations $\bar{{\bf 3}}^{l}_{L}$. $\bar{{\bf 3}}^{d}_{L}$ and ${\bf 6}_L$ are cancelled by those of the right-handed representations which are graphically shown in Figs.~(\ref{eM},\ref{dm3},\ref{uMnew}). They are denoted as  $\bar{{\bf 3}}^{l,M}_{R}$,  $\bar{{\bf 3}}^{d,M}_{R}$ and ${\bf 6}^{M}_R$ respectively. 
\begin{figure}
\begin{center}
\includegraphics[width=0.2\textwidth]{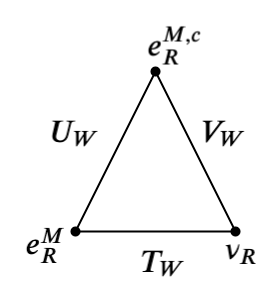}
\end{center} 
\caption{Representation $\bar{{\bf 3}}^{l,M}_{R}$}
\label{eM}
\end{figure}

\begin{figure}
\begin{center}
\includegraphics[width=0.2\textwidth]{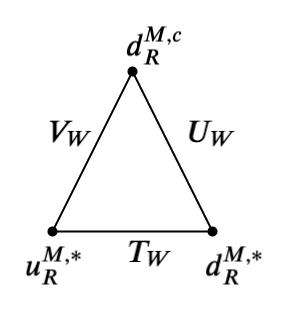}
\end{center} 
\caption{Representation $\bar{{\bf 3}}^{d,M}_{R}$}
\label{dm3}
\end{figure}

\begin{figure}
\begin{center}
\includegraphics[width=0.35\textwidth]{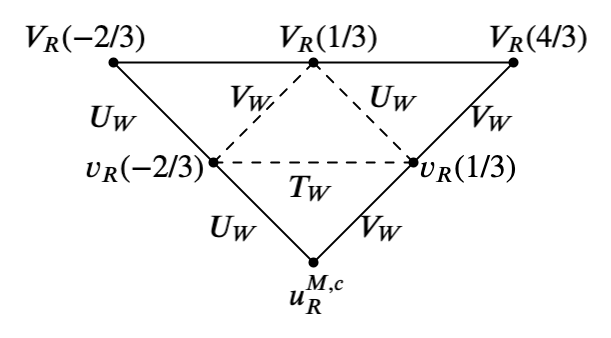}
\caption{Representation ${\bf 6}^{M}_R$}
\label{uMnew}
\end{center} 
\end{figure}

The explicit matrix forms for $\bar{{\bf 3}}^{l,M}_{L}$ and $\bar{{\bf 3}}^{d,M}_{L}$ are similar to Eqs.~(\ref{lepton},\ref{down}) with a superscript $M$ added to the fields. 
\be
\label{leptonM}
 \bar{{\bf 3}}^{M,l}_{R}= \left(
	  \begin{array}{c}
	   \nu_R (0)\\
	   e^{M}_R  (-1)\\
	   e^{M,c}_R (+1)\\
	  \end{array}
	 \right) \,.
\ee
and
\be
\label{downM}
\bar{{\bf 3}}^{M,d}_{R}= \left(
	  \begin{array}{c}
	   d^{M,*}_R (1/3) \\
	   -u^{M,*}_R (-2/3) \\
	   d^{M,c}_R (1/3)\\
	  \end{array}
	 \right) \,.
\ee
Now the total anomaly is zero, i.e. $A(\bar{{\bf 3}}^{l}_{L}+\bar{{\bf 3}}^{d}_{L}+{\bf 6}_L +\bar{{\bf 3}}^{l,M}_{R}+\bar{{\bf 3}}^{d,M}_{R}+{\bf 6}^{M}_R)=0$.

Again, let us notice that, under $SU(3)_c \times SU(3)_W$, $\bar{{\bf 3}}^{l,M}_{R}= (1, \bar{3})$, $bar{{\bf 3}}^{d,M}_{R}= (\bar{3}, \bar{3}$, and ${\bf 6}^{M}_R= (\bar{3},6)$.

From the above discussion, we can see that the mirror fermions of the EW-$\nu_R$ model which were introduced by hand in order to construct a model of non-sterile, electroweak-scale right-handed neutrinos, now appear naturally from the requirement of anomaly-freedom of the embedding group $SU(3)_W$. The seesaw mechanism with non-sterile, electroweak-scale right-handed neutrinos within our unification framework appear as a consequence rather than as something which is imposed by hand.

So far, we have been focusing on the SM and mirror fermions. In our construction of the fermion representations of $SU(3)_W$, we have discovered the singular place occupied by $u_R$ (or $u^{c}_L$): it belongs to a {\em sextet} of $SU(3)_W$ because of the value of its $U(1)_Y$ quantum number. This sextet contains new {\em vector-like quarks} (in addition to mirror fermions of the EW-$\nu_R$ model): a doublet  $ ( \upsilon_{L,R} (-2/3), \upsilon_{L,R} (1/3))$ and a triplet $(V_{L,R} (-2/3), V_{L,R} (1/3), V_{L,R} (4/3))$. We shall come back to these particles below.
\\
\\
IV) {\bf The scalar sector}
\\
\\
a) {\bf A prelude}
\\
\\
In this section, we will write down the scalar fields which are needed to give masses to the gauge bosons and to the fermions.
To prepare not just for the seesaw mechanism which involves a Dirac mass term and a Majorana mass term but also for mass terms of charged fermions, it is useful to recall some of the notations used in $SU(5)$.

A Dirac mass term in the SM involves a product such as $\bar{\psi}_R \, \psi_L$. Since all of the ingredients are grouped either in left-handed or right-handed multiplets, one uses the following identities: $\psi^{c}_{L,R}= C \bar{\psi}^{T}_{R,L}$ , $\bar{\psi^{c}}_{L,R} = \psi^{T}_{R,L} C$, where $C=\imath \gamma^2 \gamma^0$. A Dirac mass term would look like
\be
\label{dirac}
\bar{\psi^{c}}_{R} \psi_L + H.c.=  \psi^{T}_{L} C \psi_L + H.c. \,.
\ee
A Majorana mass term (for right-handed fermions) would look like
\be
\label{majorana}
\psi^{T}_{R} \sigma_2 \psi_R \,.
\ee

Some products of representations will be useful for our purpose: $3 \times 3= \bar{3} + 6$, $\bar{3} \times \bar{3}= 3 + \bar{6}$, $6 \times 6= 15+ 15+ \bar{6}$, $ 3 \times 6 = 10 +8$ and $\bar{3} \times 3 = 1 + 8$. With this, we can now list all the Yukawa couplings that give rise to the mass terms for quarks and leptons (SM + Mirror).

Both right-handed charged leptons and right-handed down quarks are accompanied by their respective doublets in the $\bar{3}$ representations. So a Dirac mass term (\ref{dirac}) can be formed by looking at $\bar{3} \times \bar{3}= 3 + \bar{6}$ for both SM and mirror fermions. A similar product is present for the Majorana mass term (\ref{majorana}). It turns out below that one needs a \{$\bar{3}$\}-Higgs for the charged lepton and down-quark Dirac mass and a \{$6$\}-Higgs for the Majorana mass.

Since $u_R$ ($u^{c}_L$) and $u^{M}_L$ ($u^{M,c}_R$) belong to $6_L$ and $6_R$ representations respectively, a Dirac mass term for the up-sector would come from the product $ 3 \times 6=10 +8$. We need a complex representation for reasons related to the global symmetry phase mentioned earlier. The appropriate Higgs representation would be a \{$10$\}-Higgs. This is also where one notices that it is unavoidable that up-quark and down-quark sectors couple to {\em two different} $SU(2)_W$ doublets (one in {\bf 3} and another one in the aforementioned {\bf 10}) since $u^{c}_L$ and $d^{c}_L$ belong to two different $SU(3)_W$ representations. This is reminiscent of the two-Higgs doublet model (2HDM) of Type II \cite{sher} where the absence of flavor-changing neutral current (FCNC) is guaranteed at tree level by imposing a discrete symmetry. Here, the need to impose such a discrete symmetry is absent for for the aforementioned reasons. 

We now turn to the classification of various Higgs fields.
\\
\\
b) {\bf Scalars in the adjoint representation}
\\
\\
We have already introduced the Higgs field, ${\bf \Phi_8=(\xi^{+}, \xi^{0}, \xi^{-}) + (\phi_X^{+},\phi^{0}_X) + (\phi_X^{-},\bar{\phi}^{0}_X) + \eta_8}$, that breaks $SU(3)_W$ down to $SU(2)_W \times U(1)_Y$, leaving two physical scalar representations under the electroweak gauge group: the {\em real triplet ${\bf \xi}$} and the {\em real singlet} ${\bf \eta_8}$. This octet Higgs field does not couple to fermions as expected. 
\\
\\
c) {\bf Scalars in fundamental representations}
\\
\\
The next step is to introduce Higgs fields that contain those of the EW-$\nu_R$ model which give masses to fermions, namely the doublets: $\Phi^{SM}_{1,2}(Y/2=\mp 1/2)=( \phi_{1}^{-} , \phi_{1}^{0}); (\phi_{2}^{0} ,  \phi_{2}^{+} )$, $\Phi^{M}_{1,2}(Y/2=\mp 1/2)=( \phi_{1}^{M,-} , \phi_{1}^{M,0}); (\phi_{2}^{M,0} ,  \phi_{2}^{M,+} )$. 
For the charged lepton and down-quark sector, we have mentioned above that the appropriate scalar is a \{$\bar{3}$\}-Higgs, one for the SM sector and one for the mirror fermions. They should contain $\Phi^{SM,M}_2$ as written down in Eq.~(\ref{massq},\ref{massl}).

We have the following fields.

\be
\label{scalar3SM}
{\bf \Phi}_{2}({\bf 3})= \left(
	  \begin{array}{c}
	   \phi_{2}^{+} \\
	   \phi_{2}^{0} \\
	   \tilde{\phi}_{2}^{-}\\
	  \end{array}
	 \right) \,,
\ee
\be
\label{scalar3M}
{\bf \Phi}^{M}_{2}({\bf 3})= \left(
	  \begin{array}{c}
	   \phi_{2}^{M,+} \\
	   \phi_{2}^{M,0} \\
	   \tilde{\phi}_{2}^{M,-}\\
	  \end{array}
	 \right) \,,
\ee
\\
\\
d) {\bf Scalars in sextet representation}
\\
\\
\be
\label{sextet}
{\bf \Phi}_6 = [(\chi^{0}, \chi^{+}, \chi^{++}), (\tilde{\phi}_{6}^{-}, \tilde{\phi}_{6}^{0}), \tilde{\phi}_{6}^{--} ] \,.
\ee

This sextet Higgs field contains the complex triplet $\tilde{\chi}= (\chi^{0}, \chi^{+}, \chi^{++})$ which is responsible for the right-handed neutrino Majorana masses.
\\
\\
e) {\bf Scalars in a decuplet representation}
\\
\\
As discussed above,  a \{$10$\}-Higgs is needed to give masses to the up-quark sector. We have
\be
\label{decupletSM}
{\bf \Phi}_{10} =[(\delta^{-}, \delta^{0}, \delta^{+}, \delta^{++}), (\sigma^{-}, \sigma^{0}, \sigma^{+}), (\phi^{-}_{1}, \phi^{0}_{1}), \omega^{-}] \,,
\ee
which couples to the SM up-quark sector, and
\be
\label{decupletM}
{\bf \Phi}^{M}_{10} =[(\delta^{-}_{M}, \delta^{0}_{M}, \delta^{+}_{M}, \delta^{++}_{M}), (\sigma^{-}_{M}, \sigma^{0}_{M}, \sigma^{+}_{M}), (\phi^{M,-}_{1}, \phi^{M,0}_{1}), \omega^{-}_{M}] \,,
\ee
for the mirror sector. Except for $\phi_{1}$ which follows the notation of the EW-$\nu_R$ model, the notations for the other fields in ${\bf \Phi}_{10}$ and ${\bf \Phi}^{M}_{10}$ are inspired by those of the famous decuplet of flavor $SU(3)$.

Notice we have {\em two real} triplets here: $(\sigma^{-}, \sigma^{0}, \sigma^{+})$ and $(\sigma^{-}_{M}, \sigma^{0}_{M}, \sigma^{+}_{M})$. Again, the question arises as to whether any one of them could play the role of the real triplet of the EW-$\nu_R$ model. This issue will be dealt with in a separate paper.   
\\
\\
f) {\bf Complex singlet scalar}
\\
\\
\be
\label{singlet2}
{\bf \phi}_S =({\bf 0})=(1,0) \,.
\ee
Notice that ${\bf \phi}_S$ refers to a singlet under $SU(3)_W$ which is automatically singlet under $SU(2)_W$ and carrying zero hypercharge. It is chosen to be so because the neutrino Dirac mass term involving the product $\bar{l}_L l_R^{M}$ would transform as $3 \times \bar{3}=1 + 8$. The octet Higgs $\Phi_8$ which breaks the symmetry, does not couple to fermions and its singlet member (real) $\eta_8$ cannot be identified with $\phi_S$ which transforms under the global symmetry discussed in the summary section. 
\\
\\
V) {\bf Yukawa interactions}
\\
\\

To write down the Yukawa interactions, we will first look at the transformation of the fermion and scalar representations under the global symmetry $U(1)_{SM} \times U(1)_{MF}$ as discussed in (\ref{SMtransformation}).
\\
\\
a) {\bf Transformations under the global symmetry $U(1)_{SM} \times U(1)_{MF}$}
\\
\\
To be consistent with the global transformations of (\ref{singlet},\ref{tripletchi}), we introduce the following transformations on the $SU(3)_W$ multiplets
\be
\label{phase1}
\bar{{\bf 3}}^{l}_{L},   \bar{{\bf 3}}^{d}_{L} \rightarrow e^{\imath \alpha_{SM}} \,  \bar{{\bf 3}}^{l}_{L}, \bar{{\bf 3}}^{d}_{L} \,,	
\ee
 \be
  \bar{{\bf 3}}^{M,l}_{R}, \bar{{\bf 3}}^{M,d}_{R} \rightarrow e^{-\imath \alpha_{MF}} \,  \bar{{\bf 3}}^{M,l}_{R}, \bar{{\bf 3}}^{M,d}_{R}
 \ee
\be
\label{phase2}
 {\bf 6}_{L}, {\bf 6}^{M}_{R} \rightarrow {\bf 6}_{L}, {\bf 6}^{M}_{R}  \,,
\ee
\be
{\bf \Phi}_{10} \rightarrow e^{-\imath \alpha_{SM}} \, {\bf \Phi}_{10} \,; \, {\bf \Phi}_{2}({\bf 3}) \rightarrow e^{2\imath \alpha_{SM}} \, {\bf \Phi}_{2}({\bf 3}) \,,
\ee
\be
\label{phase3}
{\bf \Phi}^{M}_{10} \rightarrow e^{\imath \alpha_{MF}} \, {\bf \Phi}^{M}_{10} \,; \, {\bf \Phi}^{M}_{2}({\bf 3}) \rightarrow e^{-2\imath \alpha_{MF}} \,{\bf \Phi}^{M}_{2}({\bf 3})
\ee
\be
\label{phase4}
{\bf \Phi}_6 \rightarrow e^{2\imath \alpha_{MF}} \, {\bf \Phi}_6 
\ee
\be
\label{phase5}
\phi_S \rightarrow e^{\imath(\alpha_{SM} + \alpha_{MF})} \, \phi_S
\ee

Notice that, for convenience, $u_R$ and $u^{M}_L$ have a non-zero phase transformation as written in (\ref{singlet}) but this is, in fact, unnecessary because one can redefine the transformation on  $\Phi_{10}$ and $\Phi^{M}_{10}$.. In fact, within the context of $SU(3)_W$, it is necessary to have $u_R$ and $u^{M}_L$ invariant under the global transformation because of the presence of vector-like quarks and the requirement that they get an arbitrary mass not coming from symmetry breaking.
\\
\\
b) {\bf Charged fermion mass terms}
\\
\\
\bea
\label{massq2}
{\cal L}_q &=& G_{u} {\bf 3}^{d,T}_{L} C \, {\bf \Phi}^{*}_{10} \,  {\bf 6}_{L}  + G_{d} \bar{{\bf 3}}^{d,T}_{L} C \, {\bf \Phi}^{*}_{2}({\bf 3}) \bar{{\bf 3}}^{d}_{L} +  \nonumber \\
&&G^{M}_{u} {\bf 3}^{M,d,T}_{R} C \, {\bf \Phi}^{M,*}_{10} \,  {\bf 6}^{M}_{R}  + G^{M}_{d} \bar{{\bf 3}}^{M,d,T}_{R} C \, {\bf \Phi}^{M,*}_{2}({\bf 3}) \bar{{\bf 3}}^{M,d}_{R}  \nonumber \\
&&+H.c. \,,
\eea
\bea
\label{massl2}
{\cal L}_l &=& G_{l}  \bar{{\bf 3}}^{l,T}_{L} C \, {\bf \Phi}^{*}_{2}({\bf 3}) \, \bar{{\bf 3}}^{l}_{L}  + G^{M}_{l}  \bar{{\bf 3}}^{M,l,T}_{R} C \, {\bf \Phi}^{M,*}_{2}({\bf 3}) \, \bar{{\bf 3}}^{M,l}_{R} \nonumber \\
&& + H.c. \,.
\eea
Eq.~(\ref{massq2},\ref{massl2}) give rise to the Yukawa interactions (\ref{massq},\ref{massl}).
\\
\\
c) {\bf Vector-like quark mass term}
\\
\\
Vector-like quarks obtain a gauge-invariant {\em arbitrary} mass term as follows

\be
\label{uncon}
\tilde{{\cal L}} = \tilde{M} \, {\bf \bar{6}}_{L} \, {\bf 6}^{M}_{R} + H.c. 
\ee

It might not be unreasonable to assume that the mass of the vector-like quarks $\tilde{M} \sim O(TeV)$. We shall briefly discuss at the end the implication of the mixing between the SM and mirror up-quarks.
\\
\\
c) {\bf Majorana mass term for right-handed neutrinos}
\\
\\
Following Eq.~(\ref{Majorana}), we can write the Yukawa interaction term which contains the right-handed neutrino Majorana mass term as follows
\be
\label{majorana2}
{\cal L}_M = G_M \,  \bar{{\bf 3}}^{M,l,T}_{R} \sigma_2 (\tau_2) {\bf \Phi}_{6} \bar{{\bf 3}}^{M,l}_{R} \,,
\ee
where the convention for $\sigma_2$ and $\tau_2$ follows that of Eq.~(\ref{Majorana}).
\\
\\
d) {\bf Neutrino Dirac mass term}
\\
\\
Following Eq.~(\ref{Dirac}), one can write the neutrino Dirac mass term as follows
\be
\label{dirac2}
{\mathcal L}_S = -G_{Sl} \, {\bf 3}^{l}_{L} \, \phi_S \, \bar{{\bf 3}}^{M,l}_{R}  + H.c. \,.
\ee

As discussed in \cite{pqnur}, this mixing between the SM and mirror sectors also applies to the quark sector with a similar form and $G_{Sl} \rightarrow G_{Sq}$.

\section{Implications of $SU(3)_W$ unification}

We would like to summarize the salient results of the previous sections. 1) By embedding $SU(2)_W \times U(1)_Y$ into $SU(3)_W$, $\sin^2 \theta_W$ is predicted to be $1/4$ at an energy scale $~2-3 \tev$ which is evolved down to $~0.231$ at $M_Z$ \cite{EHM}. 2) The requirement of anomaly freedom for $SU(3)_W$ gives rise to the emergence of mirror fermions. This, in turns, suggests that the seesaw mechanism is realized with non-sterile right-handed neutrinos with masses proportional to the electroweak scale just as proposed in the EW-$\nu_R$ model. 3) All the Higgs fields of  the EW-$\nu_R$ model are present in this unification. In particular, the {\em real} triplet $\xi$ appears naturally as a member of the octet Higgs which breaks $SU(3)_W$, as well as in the {\bf 10}-dimensional representations. As discussed in the Summary and electroweak monopole sections, this real Higgs triplet was needed to ensure the successful relationship $M_W = M_Z \cos \theta_W$ with an additional bonus: the possible existence of an electroweak monopole. Another surprise coming from this monopole is the prediction of $\sin^2 \theta_W$ by the imposition of the Dirac Quantization Condition (DQC) \cite{EHM}. With the $SU(3)_W$ unification, the DQC is automatically satisfied! 4) The construction of $SU(3)_W$ fermion representations leads to the unavoidable introduction of vector-like quarks, some of which come with unconventional electric charges: $ ( \upsilon_{L,R} (-2/3), \upsilon_{L,R} (1/3))$ and a triplet $(V_{L,R} (-2/3), V_{L,R} (1/3), V_{L,R} (4/3))$. 5) As seen above, the scalar sector also contains Higgs fields which are outside the framework of the EW-$\nu_R$ model.

It is beyond the scope of the paper to investigate all the implications coming out of $SU(3)_W$. We will list a few phenomena whose details will be presented elsewhere. First, we will comment on the status of experimental searches for vector-like quarks (VLQ being the popular acronym). Second, we will comment on the interactions changing $SU(2)_W$ doublet into singlet fermions mediated by the $X^{\pm}, X^{0}$ heavy gauge bosons shown in Eq.~(\ref{gauge}). 

  \subsection{I) Vector-like quarks: ${\bf V}=(V_{L,R} (-2/3), V_{L,R} (1/3), V_{L,R} (4/3))$ and {\bf v}= $( \upsilon_{L,R} (-2/3), \upsilon_{L,R} (1/3))$}

 The subject of vector-like quarks and their experimental searches is very extensive and it is beyond the scope of this paper to do a proper review. Most searches for VLQs focused on models where they come in $SU(2)_L$ doublets and/or triplets. In particular, decay signatures that are looked for are generically of the types $Q \rightarrow qW, qZ, qH$ where $Q$ is a VLQ and $q$, a SM quark. The CMS collaboration \cite{cms1} assumed a 100\% branching ratio for $T \rightarrow tZ$ and $B \rightarrow bZ$ and has put lower bounds of $1280 \gev$ and $1130 \gev$ for $T$ and $B$ respectively. The latest result from CMS \cite{cms2} searching for B-decaying  100\% into $bH$ or $bZ$ gives $1570 \gev$ and $1390 \gev$ respectively, These bounds however are not applicable to the VLQs of our $SU(3)_W$ model for the reasons given below. For a recent discussion of the sensitivity of LHC searches for VLQs, see \cite{butterworth}.
 \\
 \\
 a) {\bf Decays of $SU(3)_W$ VLQs $( \upsilon_{L,R} (-2/3), \upsilon_{L,R} (1/3))$}
 \\
 \\
 To see why the present experimental bounds on VLQs do not apply here, it is useful to go back to the graphical illustrations of the $SU(3)_W$ representations for the up-quarks: (\ref{unew}) and (\ref{uMnew}). The expression given in (\ref{TUVgauge}) indicates the gauge bosons that connect different fermions of a given representation. We now look at tree-level gauge connections. Let us look first at (\ref{unew}). Only the VLQs {\bf v}= $( \upsilon_{L,R} (-2/3), \upsilon_{L,R} (1/3))$ are connected to the SM right-handed up-quarks $u^{c}_L$ (or $u_R$) by the {\em heavy} gauge bosons $X^{\pm}_{\mu}$ ($V_W$) and $ \barparen{X^{0}}_{\mu}$ ($U_W$). The other unconventionally-charged VLQs ${\bf V}=(V_{L,R} (-2/3), V_{L,R} (1/3), V_{L,R} (4/3))$ are {\bf only} connected to {\bf v}= $( \upsilon_{L,R} (-2/3), \upsilon_{L,R} (1/3))$ by the same gauge bosons. This means that there is {\bf no tree-level} electroweak $W$ or $Z$ transition from VLQs to SM quarks. The same conclusion goes for the gauge transitions from VLQs to the mirror up-quarks as can be seen from (\ref{uMnew}).
 
Although details of a phenomenological study of the production and decays of VLQs in the $SU(3)_W$ model will be presented in a separate publication, a few estimates are in order to see where one is heading. To make an estimate, we will assume that VLQs are lighter than the gauge bosons $X^{\pm}_{\mu}$ and $\barparen{X^{0}}_{\mu}$. Let us look at the process $\upsilon(1/3) \rightarrow u^{c}_L + X^{+} \rightarrow u^{c}_L + d^{c}_L + u_L$ obtained by looking at Fig.~(\ref{unew}) and Fig.~(\ref{d3}). Writing
\be
\label{G_X}
G_X = \frac{g^2}{8 M_X^2} \,,
\ee 
One obtains
\be
\label{width}
\Gamma(\upsilon(1/3)) \approx \frac{G_X^2 \, m_{\upsilon}^5}{64 \, \pi^3} \,,
\ee 
where we ignore mixing angles and include the color factor $3$ for the $d^{c}_L + u_L$ final state. For the sake of an estimation, we take $M_X = 3 \tev$ and $m_{\upsilon} = 2 \, \tev$ giving
\be
\label{estimate}
\Gamma(\upsilon(1/3)) \approx 1.6 \, \mev \,.
\ee
This corresponds to a mean lifetime of $\tau \approx 4 \times 10^{-22} \, s$ and a decay length $l \approx 1.3 \times 10^{-13} m$. Let us notice that a typical hadronization length is $L_{h} \approx 1-10 \times 10^{-15}  m$. The decay length of the above example is much larger than the hadronization length and one would expect color-singlet hybrid (VLQ and the rest SM quarks) mesons and baryons of the types ${\cal M}_\upsilon =(\upsilon (1/3) q)$ and $\bar{{\cal M}}_\upsilon =(\bar{\upsilon}(-1/3) \bar{q})$, or ${\cal B}_\upsilon= (\upsilon (1/3) \bar{q} \bar{q})$ and $\bar{{\cal B}}_\upsilon= (\bar{\upsilon} (-1/3) q q)$ to be formed after $\upsilon(1/3) \, \bar{\upsilon}(-1/3)$ are produced in the gluon fusion process in $p \, p$ collisions at a hadron collider such as the LHC. Details such as the production cross sections and subsequent decays of the hybrid mesons and baryons will be presented in a separate paper. 

{\bf In classifying the hybrid mesons and baryons, one should keep in mind that, under color $SU(3)_c$, $\upsilon$ transforms as {\bf $\bar{3}$} while the light quarks $q$ transform as {\bf 3} and, as a result, hybrid mesons and baryons are ($\upsilon \, q$) and ($\upsilon \, \bar{q} \, \bar{q}$) bound states respectively.}
\\
\\
b) {\bf Some properties of hybrid mesons and baryons: ${\cal M}_\upsilon$ and ${\cal B}_\upsilon$}
\\
\\
The electric charge of the hybrid mesons ${\cal M}_{\upsilon(1/3),u} =(\upsilon (1/3) u)$ is $Q=+1$ and that of the meson ${\cal M}_{\upsilon(1/3),d} =(\upsilon (1/3) d)$ is $Q=0$. Similarly, one has ${\cal M}_{\upsilon(-2/3),u} =(\upsilon (-2/3) u)$ ($Q=0$) and ${\cal M}_{\upsilon(-2/3),d} =(\upsilon (-2/3) d)$ ($Q=-1$). Those of the hybrid baryons ${\cal B}_{\upsilon(1/3),\bar{u},\bar{u}}= (\upsilon (1/3) \bar{u} \bar{u})$, ${\cal B}_{\upsilon(1/3),\bar{u},\bar{d}}= (\upsilon (1/3) \bar{u} \bar{d})$, and ${\cal B}_{\upsilon,\bar{d},\bar{d}}= (\upsilon (1/3) \bar{d} \bar{d})$ are $-1$, $0$ and $+1$ respectively. Similarly, one has ${\cal B}_{\upsilon(-2/3),\bar{u},\bar{u}}= (\upsilon (-2/3) \bar{u} \bar{u})$, ${\cal B}_{\upsilon(-2/3),\bar{u},\bar{d}}= (\upsilon (-2/3) \bar{u} \bar{d})$, and ${\cal B}_{\upsilon(-2/3),\bar{d},\bar{d}}= (\upsilon (-2/3) \bar{d} \bar{d})$ with electric charges $-2$, $-1$ and $0$ respectively. 
It is easy to see that non-hybrid heavier mesons of the types $\upsilon(i) \bar{\upsilon}(j)$ and $\upsilon(i) \upsilon(j) \upsilon(k)$, where $i,j,k$ refer to the charges $-2/3$ and $1/3$, are also all {\em integrally-charged}. Of course, these non-hybrid mesons and baryons composed solely of the VLQ $( \upsilon_{L,R} (-2/3), \upsilon_{L,R} (1/3))$ are out of reach of the LHC. The hybrid ones are however accessible at the LHC and it would be interesting to study in details their signatures, a topic to be covered in the follow-up paper. The above hybrid mesons and baryons are summarized in the tables below.

\begin{table}[h!]
\begin{center}
\caption{Hybrid mesons ${\cal M}_{\upsilon,\bar{q}}$}
\label{hybridmeson1}
\begin{tabular}{|l|c|c||c|r|}
\hline
 &${\cal M}_{\upsilon(-2/3),u}$&${\cal M}_{\upsilon(-2/3),d}$& ${\cal M}_{\upsilon(1/3),u}$&${\cal M}_{\upsilon(1/3),d}$  \\
 \hline
Q= &0&-1&+1&0 \\
\hline
\hline
 \end{tabular}
\end{center}
\end{table}
\begin{table}[h!]
\begin{center}
\caption{Hybrid baryons ${\cal M}_{\upsilon,q,q}$}
\label{hybridbaryon1}
\begin{tabular}{|l|c|c|r|}
\hline
 &${\cal B}_{\upsilon(-2/3),\bar{u},\bar{u}}$&${\cal B}_{\upsilon(-2/3),\bar{u},\bar{d}}$&${\cal B}_{\upsilon(-2/3),\bar{d},\bar{d}}$\\
 \hline
Q= &-2&-1&0 \\
\hline
\hline
 &${\cal B}_{\upsilon(1/3),\bar{u},\bar{u}}$&${\cal B}_{\upsilon(1/3),\bar{u},\bar{d}}$&${\cal B}_{\upsilon(1/3),\bar{d},\bar{d}}$\\
 \hline
Q= &-1&0&+1 \\
\hline
\hline
 \end{tabular}
\end{center}
\end{table}
c) {\bf Unconventionally-charged VLQ ${\bf V}=(V_{L,R} (-2/3), V_{L,R} (1/3), V_{L,R} (4/3))$}
\\
\\
Fig.~(\ref{unew}) shows how ${\bf V}=(V_{L,R} (-2/3), V_{L,R} (1/3), V_{L,R} (4/3))$ couples to the other quarks in the sextet ${\bf 6}_L$. We see that they couple directly only to $( \upsilon_{L,R} (-2/3), \upsilon_{L,R} (1/3))$ by the $X^{\pm,0}$ gauge bosons. However, we see from Eq.~(\ref{uncon}) that the triplet and doublet VLQs are {\em degenerate in mass} if that is the only source. 

Unless we can break that degeneracy, ${\bf V}=(V_{L,R} (-2/3), V_{L,R} (1/3), V_{L,R} (4/3))$ cannot decay into $( \upsilon_{L,R} (-2/3), \upsilon_{L,R} (1/3))$ and would be stable.The hybrid mesons and baryons formed from these quarks with the SM quarks are interesting objects to search for. As with the previous case, the non-hybrid mesons and baryons would be too heavy and would not be accessible at the LHC. We will present various scenarios for these hybrid hadrons in a subsequent paper. For the moment, we simply list the kinds of hybrid mesons and baryons of interest. {\bf Again, one should keep in mind that $V$ transforms, under $SU(3)_c$ , as {\bf $\bar{3}$} while the light quarks $q$ transform as {\bf 3}.}

Here we have: 1) ${\cal M}_{V(-2/3),u} =(V(-2/3) u)$ ($Q=0$), ${\cal M}_{V(-2/3),d} =(V(-2/3) d)$ ($Q=-1$); 2) ${\cal M}_{V(1/3),u} =(V(1/3) u)$ ($Q=+1$), ${\cal M}_{V(1/3),d} =(V(1/3) d)$ ($Q=0$); 3) ${\cal M}_{V(4/3),u} =(V(4/3) u)$ ($Q=2$), ${\cal M}_{V(4/3),d} =(V(4/3) d)$ ($Q=+1$) for the hybrid mesons. For the hybrid baryons, we have: 1) ${\cal B}_{V,\bar{u},\bar{u}}= (V(-2/3) \bar{u} \bar{u})$ ($Q=-2$), ${\cal B}_{V,\bar{u},\bar{d}}= (V(-2/3) \bar{u} \bar{d})$ ($Q=-1$), ${\cal B}_{V,\bar{d},\bar{d}}= (V(-2/3) \bar{d} \bar{d})$ ($Q=0$); 2) ${\cal B}_{V,\bar{u},\bar{u}}= (V(1/3) \bar{u} \bar{u})$ ($Q=-1$), ${\cal B}_{V,\bar{u},\bar{d}}= (V(1/3) \bar{u} \bar{d})$ ($Q=0$), ${\cal B}_{V,\bar{d},\bar{d}}= (V(1/3) \bar{d} \bar{d})$ ($Q=+1$); 3) ${\cal B}_{V,\bar{u},\bar{u}}= (V(4/3) \bar{u} \bar{u})$ ($Q=0$), ${\cal B}_{V,\bar{u},\bar{d}}= (V(4/3) \bar{u} \bar{d})$ ($Q=+1$), ${\cal B}_{V,\bar{d},\bar{d}}= (V(4/3) \bar{d} \bar{d})$ ($Q=2$).  

 These hybrid mesons and baryons are summarized in the tables below
 \begin{table}[h!]
\begin{center}
\caption{Hybrid mesons ${\cal M}_{V,\bar{q}}$}
\label{hybridmeson2}
\begin{tabular}{|l|c|r|}
\hline
 &${\cal M}_{V(-2/3),u}$&${\cal M}_{V(-2/3),d}$  \\
\hline
Q= &0&-1 \\
\hline
\hline
 &${\cal M}_{V(1/3),u}$&${\cal M}_{V(1/3),d}$ \\
\hline
Q= &+1&0 \\
\hline
\hline
 &${\cal M}_{V(4/3),u}$&${\cal M}_{V(4/3),d}$  \\
 \hline
Q= &2&+1 \\
\hline
\hline
 \end{tabular}
\end{center}
\end{table}
\begin{table}[h!]
\begin{center}
\caption{Hybrid baryons ${\cal M}_{V,q,q}$}
\label{hybridbaryon2}
\begin{tabular}{|l|c|c|r|}
\hline
 &${\cal B}_{V(-2/3),\bar{u},\bar{u}}$&${\cal B}_{V(-2/3),\bar{u},\bar{d}}$&${\cal B}_{V(-2/3),\bar{d},\bar{d}}$\\
 \hline
Q= &-2&-1&0 \\
\hline
\hline
 &${\cal B}_{V(1/3),\bar{u},\bar{u}}$&${\cal B}_{V(1/3),\bar{u},\bar{d}}$&${\cal B}_{V(1/3),\bar{d},\bar{d}}$\\
 \hline
Q= &-1&0&+1 \\
\hline
\hline
 &${\cal B}_{V(4/3),\bar{u},\bar{u}}$&${\cal B}_{V(4/3),\bar{u},\bar{d}}$&${\cal B}_{V(4/3),\bar{d},\bar{d}}$\\
 \hline
Q= &0&+1&+2 \\
\hline
\hline
 \end{tabular}
\end{center}
\end{table}

At this point, it is important to stress the most interesting result of this section: the presence of a {\bf doubly-charged meson} ${\cal M}_{V(4/3),u}$ ($Q=\pm 2 e$). This is possible because of the existence of a charge $4/3$ vector-like quarks $V(4/3)$. SM quarks with electric charges $\pm 2/3$ and $\pm 1/3$ can only accommodate mesons with a maximum electric charge $1$: There is {\em no} ordinary meson with charge $2e$. (Ordinary baryons having a charge $Q=2 e$ such as $\Delta^{++}$ are expected.)
\\
\\
II) {\bf Comments on interactions mediated by the heavy $X^{\pm,0}$ gauge bosons}
\\
\\
The heavy $X^{\pm,0}$ gauge bosons mediate the interactions between an $SU(2)_W$ doublet and a singlet, for instance between $(\nu_L, e_L)$ and $e^{c}_L$ (or $e_R$) as one can see from Fig.~(\ref{e3}). Such interactions are expected to be much weaker than the SM weak interactions. At first sight, one might be tempted to see whether limits on the masses of these gauge bosons have already been set from searches for sequential $W$s in processes such as $W^{\prime} \rightarrow \nu e$ ($6 \, \tev$ from ATLAS \cite{rpp}). However, it is fair to say that a detailed analysis for the present model is needed in order to see whether or not those bounds are applicable. One could also  look for them in precision experiments, for example. The processes which could reveal signatures of such interactions are under study. One possibility is the asymmetry in polarized $e^{+}\, e^{-}$ scattering which will involve, beside the dominant Z-boson exchange, a subdominant $X^{0}$ exchange. Another place where one might be able to measure a left-right asymmetry in polarized electron scattering on unpolarized proton is the $Q_{weak}$ collaboration experiment at Jefferson Lab.

\section{Conclusion}

The EW-$\nu_R$ model  containing a real Higgs triplet was found to accommodate a topologically stable, finite-energy electroweak monopole \cite{pq_monopole}. A subsequent important and unexpected observation was made concerning this electroweak monopole \cite{EHM}. By imposing the Dirac Quantization Condition on this monopole, \cite{EHM} made a prediction for $\sin^2 \theta_W$ to be equal to {\bf 1/4} at the mass scale $M_M$ of the monopole ($2-3 \tev$). This value of $1/4$ was evolved down at one loop to $\sim 0.231$ at $M_Z$, in agreement with experiment. Now, ${\sin}^2\theta_W(M_M^2) = g'^2/[g^2+g'^2]=1/4$ implies $\alpha^\prime =(1/3) \alpha_2$ ($g$ and $g^{\prime}$ are the $SU(2)_W$ and $U(1)_Y$ couplings respectively). This fact urges us to ask whether or not $U(1)_Y$ could be unified with $SU(2)_W$ in a consistent way in a larger group which is free of anomalies and which contains all the degrees of freedom of the EW-$\nu_R$ model. 

The answer to the above question is affirmative. It is shown above that the simplest group that embeds $SU(2)_W \times U(1)_Y$ and which yields $\sin^2 \theta_W=1/4$ is $SU(3)_W$. We call this the {\em weak eightfold way}. The following results are obtained. 
\\
\bi
\item $SU(3)_W \rightarrow SU(2)_W \times U(1)_Y$ by a Higgs octet. With the $SU(3)_W$ generator $T_8$ identified as the $U(1)_Y$ generator, $\sin^2 \theta_W=1/4$ is obtained straightforwardly. 

\item  The SM lepton doublets $l_L$ and singlets $e_R$ (written as left-handed charge conjugates $e^{c}_L$) can be assembled into {\bf $\bar{3}$}. The same goes for SM quark doublets $q_L$ and singlets $d_R$ (written as $d^{c}_L$). This would leave $u_R$ (or $u^{c}_L$) as an "orphan". It is found that, in order to put $u_R$ in an $SU(3)_W$ multiplet to complete the assignment, one would need a sextet ({\bf 6}) which includes, beside $u^{c}_L$, a $SU(2)_W$ left-handed doublet with conventional electric charge, $( \upsilon_{L} (-2/3), \upsilon_{L} (1/3))$, and an $SU(2)_W$ left-handed {\em triplet} with unconventional electric charges, $(V_{L} (-2/3), V_{L} (1/3), V_{L} (4/3))$. As discussed in the text, $SU(3)_W$ {\em would not} be anomaly-free if the aforementioned representations were the only ones present. In addition, the extra quarks would be massless in the absence of their right-handed counterparts (or equivalently $f^{c}_L$). The fact that $d^{c}_L$ and $u^{c}_L$ belong to different representations is not surprising as we have seen this pattern in constructions of Grand Unified models. For instance, in $SU(5)$, $u^{c}_L$ belongs to a {\bf 10} while $d^{c}_L$ belongs to a {\bf $\bar{5}$}.

\item  It is shown that the only option for making $SU(3)_W$ anomaly-free is to have matching right-handed representations, giving rise to the so-called mirror fermions of the EW-$\nu_R$ model. Turning the argument around, one might imagine embedding $SU(2)_W \times U(1)_Y$ into $SU(3)_W$ first to obtain $\sin^2 \theta_W=1/4$ (followed by its evolution from a mass scale $\sim 3 \tev$ down to $M_Z$). One would discover that one needs mirror fermions to make $SU(3)_W$ anomaly-free which, in turn, leads to the model of non-sterile, electroweak-scale right-handed neutrinos. These mirror right-handed representations contain matching right-handed $( \upsilon_{R} (-2/3), \upsilon_{R} (1/3))$ and $(V_{R} (-2/3), V_{R} (1/3), V_{R} (4/3))$ transforming these extra quarks into Vector-Like Quarks (VLQ).  

\item  The scalar sector is constructed and contains the Higgs fields present in the EW-$\nu_R$ model. Because $u^{c}_L$ and $d^{c}_L$ belong to {\em two different} $SU(3)_W$ representations, it is found that thee up-quark and down-quark sectors couple to two different scalar representations {\bf 10} and {\bf 3} respectively and, hence, to two different $SU(2)_W$ Higgs doublets. The absence of FCNC at tree level is automatically guaranteed. The right-handed Majorana mass term comes from the scalar belonging to the {\bf 6}-representation which contains the complex $SU(2)_W$ triplet $\tilde{\chi}$. Real Higgs triplets are found in the octet as well as in the {\bf 10}. A complex $SU(3)_W$-singlet Higgs field is introduced which plays the role of $\phi_S$ in  the EW-$\nu_R$ model. Finally, there is a {\em real} $SU(2)_W \times U(1)_Y$ singlet (named $\eta_8$ here) which is part of the octet and which could be a DM candidate. Such a possibility is under investigation.

\item As discussed in the section on Vector-Like Quarks, the VLQs of the $SU(3)_W$ model either have a decay length exceeding the hadronization length ($\upsilon$) or are "stable ($V$). As a result, mesons and baryons can be formed. In particular, {\em hybrid} mesons and baryons composed of one VLQ and one or two ordinary quarks (meson or baryon) are very interesting objects to look for. In particular, the model predicts the existence of a {\bf doubly-charged meson} ${\cal M}_{V(4/3),u}$ ($Q=\pm 2 e$) which does not exist with ordinary quarks. An observation of such a meson would reveal the existence of quarks with {\em unconventional electric charges} such as $V(4/3)$. As we discuss in the section on fermion representations, the existence of this kind of unconventionally-charged quark is {\em inevitable} because, in order to fit the ("orphaned") right-handed up-quark ($u_R$ or $u^{c}_L$) into a representation of $SU(3)_W$, one would require a sextet, leading to such a possibility. 

\item Another interesting topic not discussed here relates to Eq.~(\ref{uncon}) which gives a gauge-invariant mass term for VLQs. It also induces in the top-quark sector a mixing between SM and mirror quarks. Implications concerning the hierarchy in the top-quark sector will be investigated.

\item Last but not least, the model contains gauge interactions mediated by heavy TeV-scale gauge bosons, linking members of a $SU(2)_W$ doublet (e.g. $(\nu_L, e_L)$) to a singlet (e.g. $e_R$ ($e^{c}_L$)). Signals of such interactions can be searched for at the LHC and in precision experiments such as the left-right asymmetry in polarized electron scattering on unpolarized proton at Jefferson Laboratory.
 \ei
 
 It goes without saying that the results of the construction of the {\em weak eightfold way} presented here along with some of the implications are preliminary. Further phenomenological consequences will be  dealt with in a separate publication. 
 
 \begin{acknowledgements}
 I would like to thanks Julian Heeck, Tc Yuan, and Eduardo Peinado for discussions.
 \end{acknowledgements}

\end{document}